\documentclass[prb,preprint]{revtex4-1} 
\usepackage{amsmath}  
\usepackage{amsfonts} 
\usepackage{graphicx} 

\begin{document}

\title{Rubidium Isotope Shift Measurement using Noisy Lasers}

\author{Theodore J. Bucci}
\email{tjbucci@student.ysu.edu} 
\author{Jonathan Feigert}
\email{jpfeigert@student.ysu.edu} 

\author{Michael Crescimanno}
\email{dcphtn@gmail.com}

\affiliation{Department of Physics and Astronomy, Youngstown State University, Youngstown, OH, 44555}

\author{Brandon Chamberlain}
\email{chamberlain.187@buckeyemail.osu.edu}
\author{Alex Giovannone}
\email{adgiovannone@gmail.com}
\affiliation{Department of Physics, The Ohio State University, Columbus, OH. }

\date{\today}

\begin{abstract} 
Data students collect from the typical advanced undergraduate laboratory on Saturated Absorption Spectroscopy (SAS) of rubidium can be used to measure the isotope shift and thus leads to an estimate of the isotopic ground state energy shift. This helps students refine their `picture' of the atomic ground state. We describe theoretically why this laboratory works well with free-running laser diodes, demonstrate it experimentally using these lasers tuned to either principal near-infrared transitions, and show an extension of the laboratory using the modulation transfer spectroscopy method. 
\end{abstract}

\maketitle 

\section{Introduction:} 

In most instances of the Saturated Absorption Spectroscopy (SAS) laboratory in current use in the advanced undergraduate laboratory, the goals are to understand the conceptual basis of saturation,  its use in a Doppler-free measurement and to determine the excited state hyperfine splittings for the two isotopes, $^{85}$Rb and $^{87}$Rb. While these are undoubtedly of great value in their own right, one can extend the value of this laboratory experience by using the SAS data to determine the isotope (energy-)shift between the two isotopes. A recent (summer 2019) survey of all Ohio colleges and universities offering BS/BA physics determined that while fully one third of the 20 programs have laser spectroscopy of the rubidium atom as a prominent part of their advanced laboratory, not one (other than  Youngstown State University (YSU)) uses it to measure the isotope shift between rubidium isotopes. Of importance later, that survey also determined that every instance of that laboratory used an External Cavity Diode Laser (ECDL) system, instead of the much simpler and less expensive "noisy" or free-running laser diode.  
 
The pedagogic value of doing an isotope shift laboratory is that it refines and expands upon the atomic physics lessons that the students learn as part of their junior/senior physics sequence. Asking undergraduate physics majors about their 'picture' of the atom often reveals common misconceptions, among them the notion that the nucleus 'sits' at the center of a symmetrical electron cloud. If so, then why would the mass of the nucleus itself play any role in the electronic excitation spectrum? Of course, an appeal to their classical understanding reveals that the nucleus must indeed be in countermotion to the electron, which in the S-state is continually moving radially, but a laboratory experience measuring the isotope shift (in part due to the countermotion) can serve to reinforce this `picture' of co-motion inside the atom. Isotope shift measurements in the optical section of the undergraduate advanced laboratory appears to have traditionally been done by comparing the hydrogen and deuterium Balmer series using a relatively low resolution spectrometer.\cite{survey2} Doing so one recovers the aforementioned effect of the nuclear orbital co-motion without being sensitive to the much smaller isotope nuclear size effects. In contrast, the experiment in rubidium contains both the motional and the nuclear size effects. 

\begin{figure}[!tbp]
\centering 
\includegraphics[width=6in]{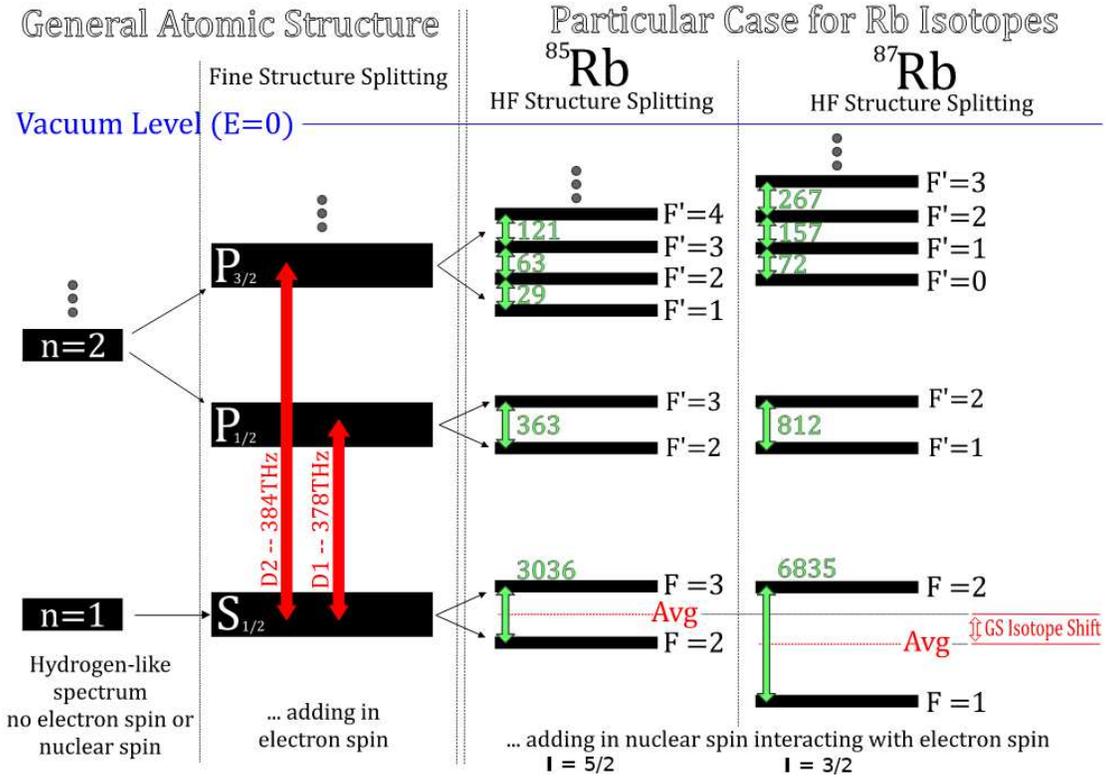}
\caption{An explanation of the level structure of single electron atoms, starting from the classical Hydrogen spectrum with spinless electrons (Schroedinger equation). As well described in undergraduate expositions of the corrections to the hydrogen spectra, next come relativistic contributions which include the spin of the electron and cause the fine structure splitting. Next order to that are nuclear magnetic effects causing the atomic levels to be organized in terms of the total angular momentum $F$, resulting in hyperfine structure. Small letters in green are the measured adjacent line splittings in MHz. This paper describes an elementary laboratory for measuring the ground state isotope shift. Not all atomic levels are shown.}
\label{levelStruct}
\end{figure}

Laser diodes have long been used in hyperfine spectroscopy on both of the principal near infrared optical transitions (called the `D1' at 795nm or `D2' at 780nm respectively). See Figure [\ref{levelStruct}]) of the rubidium atom, \cite{nez} and SAS on these transitions as reported in the extensive existing literature typically employ ECDLs,\cite{olsen, preston} or somewhat more intricate cavity stabilized schemes.\cite{barwood1, barwood2} For more details on the physics of the laser diode and the historical development of their use in atomic spectroscopy, see Ref.~\cite{comparo2}. There is significant literature on SAS's use in higher excited state levels, notably, \cite{ko, gustafsson, gustafsson2, moon, olsen} and even pulsed lasers, for example in Ref.~\cite{barwood2, prijapati}, as well as for other atomic species,\cite{pappas} even with free-running (non ECDL) laser diodes.\cite{brandenberger}

The laser diodes used for these experiments are commodity Indium Gallium Arsenide NIR diodes and have a fairly  small temperature coefficient for the emission wavelength. As is well known, the material's modest wavelength temperature coefficient results in the emission wavelength varying smoothly with temperature and current over limited ranges. These continuous variations are accompanied by reproducible discontinuous so-called ``mode hops" in the laser's wavelength (and output intensity) over larger modulations in the diode's current or temperature. Nevertheless, about half the diodes we have tested this way end up being suitable for these experiments in that they allow for a mode-hop-free sweep range that easily spans the entire rubidium D1 (or D2) transitions (roughly 9 GHz). An additional barrier to adoption parallel with the mechanical/optical complexity and cost of the ECDL laser head is also its intricate temperature stabilization circuit (see for example that in Ref.~\cite{wieman}), for which there now are cheaper commercial options.

Using D-line spectroscopy to quantify the isotope effect also has a celebrated history, \cite{hughes, bruch,roberts, arimondo, corney1, corney2} (read Ref.~\cite{aldridge} for a more up-to-date experimental summary of isotope shift measurements in rubidium) and more recently precision measurements of the isotope shift have been of interest for the indirect detection of new physical interactions.\cite{yamamoto1, yamamoto2, berengut} More modern laser diode spectroscopic approaches to measurements of the isotope shift have been done in warm vapor cells \cite{gibbs, banerjee} as well as in the experimentally more challenging discharge tubes and atomic beams.\cite{beacham,stoicheff0, stoicheff1, camparo} 

Over the last 20+ years, much of the laser diode spectroscopy technique has become part of the canon of the advanced undergraduate laboratory.\cite{blue}  Of particular interest beyond measuring the isotope shift, laser diode spectroscopy has been used to teach about Doppler broadening,\cite{leahy} radiative broadening,\cite{bali} non-linear spectroscopy, \cite{jaques} noise spectroscopy, \cite{schulte} SAS lineshape,\cite{sherlock, smith} again with these studies using ECDLs. It is noteworthy that there is literature\cite{rao} of early spectroscopic experiments using only a free-running laser diode. However this experiment did not use SAS and only discussed, but did not measure, the isotope effect in rubidium. There is as well another report\cite{wizemann} using SAS via illumination by a  free-running laser diode, but did not take the measurements needed to determine an isotope shift. 
  
The goals of this article are to describe the simplicity and pedagogic utility of the laser spectroscopic determination of the isotope shift in atomic rubidium and to remind colleagues and explain why noisy, free-running lasers work so well in this application. By "noisy" laser, we are referring to the frequency (or phase) jitter intrinsic to the laser diode's normal operation. Note that, somewhat surprisingly, the spectral widths of free running laser diodes can be as much as 5 times that of the measured width of the SAS resonances they create. This means in practice that, even if the lasers used are about 10 to 15 times spectrally broader than the natural line width of the transitions that they probe, one can still easily capture MHz-level excited state frequency differences. As noted,  most undergraduate laboratories of rubidium SAS (and others involving atomic physics methods but with warm atoms) use ECDLs, ranging in cost typically from \$5k to \$20K. We find this curious in light of the fact that, as we describe here,  much less expensive and simpler laser diode sources apparently work equally well. Below, by directly comparing the isotope shift measurements using simply collimated and temperature controlled laser diodes to that of ECDLs on both the D1 and D2 transition, we find essentially no statistical difference in accuracy. 

We provide a brief, simplified analysis of the canonical quantum optics model for the SAS process, using it to explain the weak dependence of the signal quality (resonances centers, contrast and linewidths) on the spectral width of the interrogating laser.  After that section we describe the experimental configuration, detail the data reduction step and include, as summary of all these experiments performed here at YSU last year, a tabulation of the measured isotope shifts by this method using noisy, free-running laser diodes and ECDLs on both the D1 and D2 transitions.  

\section{Theory:}
In its simplest form, isotope shifts in a spectrum are due primarily to two phenomena: the dynamical effect of the nuclear mass and electron's co-motion  and the residual differences in the electron's interaction with the nucleus (see Ref.~\cite{palffy} for a student-friendly, very readable and comprehensive current theory review of isotope shifts in complex atoms). On dimensional grounds alone but ignoring the hyperfine interactions, the energy $E$ of an atom can only be a function of the electron mass $m$ and the nuclear mass $M$.  Asymptotic analysis of the problem indicates that the relevant mass scale for the electron-nucleus interaction is the reduced mass $\mu=mM/(m+M)$, whereas for the electron-electron interaction, it is $m/2$; so that we may write for each orbital and energy  $E = \mu W(m/M, e,s) + m Y(m/M, e,s)$ where $e$ is the (dimensionless) electron charge and `$s$' represents the quantum numbers of that level and where $W$ and $Y$ are dimensionless functions. For example, the familiar Rydberg formula (hydrogen $Z=1$) reads $E_n = \frac{\mu c^2 Z^2 \alpha^2}{2n^2}$ where 1/$\alpha$ = 137.03 is the fine structure constant, $c$ is the speed of light and $n$, the principal quantum number. The part of the isotope shift that is a consequence of this overall
$\mu$-scaling of the first term in the energy in literature is referred to as ``the normal mass shift" , or NMS. The second $Y$ term is smaller and referred to as the ``nuclear shift" or NS. It is a non-nuclear mass related effect attributable in part to the difference in the mean charge radius of the isotope nuclei. This difference impacts the energies of the electrons
(particularly the $S$-electrons ($L=0$)) as they sample the difference between the exact (and thus singular) Coulomb potential and that of a `smeared' nuclear charge. Here for simplicity of discussion we include in the NS contributions from secondary changes in the multielectron wavefunction in moving between isotopes. Finally, also included in the NS, are any contributions from the magnetic properties of the nucleus beyond the hyperfine splittings which mark the (relative) orientations of the electron spin and magnetic moment of the isotope's nucleus.

Much of the additional complexity of using rubidium isotopes, as opposed to the spectral comparison between those of hydrogen, is primarily due to the isotope shift being much smaller in the former, so for rubidium isotopes one must contend with the hyperfine splittings of both the ground state and excited states. Fortunately this is simple and straightforward to do, since the hyperfine level shifts are parameterized via\cite{arimondo, corney1}
\begin{equation}
\label{hyperfineLevels}
\delta E(I,L,S) =E_{0HFS} +  \frac{A}{2}C + \frac{B}{2}\frac{\frac{3}{4}C(C+1)-J(J+1)I(I+1)}{I(2I-1)J(2J-1)}  
\end{equation} 
where $E_{0HFS}$ would be the shift in the absence of any hyperfine interaction. 
Here $C = F(F+1)-J(J+1)-I(I+1)$ and $F = I+L+S$  is the total 
atomic angular momentum, written as a sum of the spin of the nucleus, the electron orbital contribution and the electron intrinsic spin $S$ ($S=\frac{1}{2}$). Note further that the `$B$' term does not contribute for the D1 transition since here $J=\frac{1}{2}$ for both ground state and excited state ($A$ = 120.5MHz and 408.3MHz, resp. for $^{85}$Rb and $^{87}$Rb). In general, for the high-Z alkali atoms, the `$B$' term is generally significantly smaller than the `$A$' term. For the D2 transition in $^{85}$Rb accepted values are $A$ = 25.0MHz $B$ = 25.7MHz and for $^{87}$Rb $A$ = 84.7MHz, and $B$ = 12.5MHz.\cite{barwood2, corney2}
 
In what follows, we assume that students have in earlier experiments already measured the ground state hyperfine splitting and here adopt the accepted values of  $\Delta E_{HFS}/h =$ 3036 ($^{85}$Rb) MHz and 6835 ($^{87}$Rb) MHz. Simply fitting for line centers of the linear absorption features themselves in a vacuum rubidium vapor cell, and subsequently finding the ground state hyperfine `center of mass' for each isotope (by performing the weighted sum with respect to the ground state multiplicities alone, see below) would suffice for determining the isotope shift were it not for the fact that the excited state hyperfine splittings are so large they significantly distort the linear absorption features, making this direct approach untenable. We must instead resolve the individual excited state resonance line centers and then combine these data. SAS is one convenient technique for doing exactly that. 
  
Once the spectral locations for the excited state hyperfine levels are measured, we must account for the `s' (quantum number) dependent part of $W$ and $Y$. For simplicity of presentation, we largely ignore the ``$B$" term contributions to the hyperfine splitting and take the hyperfine energy splittings to be approximately $dE (s) = \frac{1}{2} A (F(F+1)-I(I+1)-J(J+1))$, for $A$ a constant proportional the nuclear magnetic moment and $F = I+J$ where $J = L+S$, in the usual notation familiar to first year quantum mechanics students. Note that for $L=0$ (ground states) the $J=\frac{1}{2}$ and so $F_{\pm}  = I \pm \frac{1}{2}$ , leading to $dE_{\pm}  = \pm \frac{A}{4}(2F_{\pm} +1)$. The usual excited state hyperfine interval then is
$\Delta E_{hfs} = dE_+ -dE_-$ =   3$A_{85}$ or  2$A_{87}$  for $^{85}$Rb and  $^{87}$Rb respectively. 

The appearance of the state multiplicities $2F_{\pm} +1$ in the hyperfine energies allows us to remove the hyperfine energy splittings entirely from the spectrum by forming  linear combination of the energies, for example for the ground states $E_{0HFS~^{85}Rb} = (7dE_++5dE_-)/12$ for $^{85}Rb$ and $E_{0HFS~^{87}Rb}=(5dE_++3dE_-)/8$  for $^{87}Rb$. 
These so-called `center of mass' energies represent the energy of the atomic state were the electron spinless and, to re-iterate, are simply the state-weighted average of the (fixed total F) state energies. Intuitively one can think of this sum as homogenizing all possible relative spin orientations of the electron and the nucleus, causing a vector interaction ({\it i.e.} the `$A$' part of the hyperfine interaction) to average to zero. We apply this same method of weighted sums (though with the excited state multiplicities) to the different excited state hyperfine intervals we measure from SAS. The hyperfine interaction independent energy level spacings that result from this multiplicity-weighted averaging are then compared between isotopes. 

The Saturated Absorption Spectroscopy (SAS) used widely in undergraduate laboratories for measurement of the $A$ and $B$ coefficients (or the grounds state hyperfine interval) results in a fine enough comparison of the $^{85}$Rb and $^{87}$Rb spectra for a determination of the isotope shift at the $\sim$10\% level with no extra experimental work.
Although there are many other ways of measuring the isotope shift, SAS is a convenient, simple, and accurate probe of the multicomponent nature of each of the absorption lines and, with its sub-Doppler linewidths, greatly simplifies the determination of the individual transitions line centers. 
There are two other, though less obvious, advantages to using SAS. The SAS line centers are somewhat protected from the usual vagaries of AC stark effects, beam mismatch, and SAS lineshape distortions that accompany the unavoidable optical pumping processes in an SAS optical field. Importantly, we show below that the SAS resonance linewidth and line contrast are not strongly perturbed by the laser's spectral width  even when it is many 10's of MHz wide, that is, many times the natural linewidth being excited. This result is consistent with the experience using an ECDLs, as Ref.~\cite{preston} summarizes with the observation that the width of SAS resonances in a vacuum vapor cell are typically as much as 6 times that of the natural linewidth. ECDL's typically have spectral widths significantly narrower than the natural width of the optical transitions in the rubidium atom themselves. 

A simplified theoretical model explains why (spectrally noisy) free-running non-ECDL laser sources suffice for  SAS-based measurements of the excited state hyperfine intervals. SAS is deeply understood \cite{hansch, holt} and although what we present here is not new, we include it for logical completeness and feel it of value because so many teachers of the advanced laboratory courses that have a laser spectroscopy of rubidium lab are surprised to learn that the SAS resonance widths are not only sub-Doppler but can be sub-laser linewidth. 

While passing through a medium, light's electric and magnetic fields are modulated by the AC polarizability of that medium. This can be most directly described in terms of the off diagonal elements (called ``coherences") of the density matrix $\rho$ whose elements are $\rho_{ij} = |i\rangle\langle j|$ where $i,j$ = ``g" for ground state and ``e" for excited state. The diagonal parts of the density matrix are proportional to the state occupations, and thus are called ``populations." For the two-level system  illuminated by a single laser field,  the density matrix elements in the rotating wave approximation, satisfy,\cite{boyd} 
\begin{equation}
    \rho_{eg} = \rho_{ge}^* =  \frac{i\Omega(\rho_{gg}-\rho_{ee})}{{\Delta^2 + \gamma^2 + 2\Omega^2}} \qquad \rho_{gg} = \frac{\Delta^2+\gamma^2+\Omega^2}{\Delta^2 + \gamma^2 + 2\Omega^2}\qquad \rho_{gg}+\rho_{ee} = 1
\label{QO1} 
\end{equation} 
where the Rabi frequency $\Omega$ is the product of the laser beam's electric field and the dipole matrix element connecting the states ``e" and ``g", $\gamma$ is the decay constant of the state ``e" and $\Delta = \omega-\omega_0$ is the frequency detuning,  the frequency mismatch between the ``e-g" transition frequency $\omega_0$ and the laser frequency $\omega$. We have suppressed subtleties of the difference between the population (i.e. $\rho_{ee}$ and $\rho_{gg}$) decay constant $\gamma$ and that of the coherences, for simplicity assuming that they are related to each other in the simplest way. The absorption of the light is proportional to $dn {\rm Im}(\rho_{eg}\Omega)$ where $n$ is the number density of the atoms and $d$ is the dipole matrix element of the transition. Note that in the high intensity limit ($\Omega \gg \gamma$) we have $\rho_{ee}= \rho_{gg} = 1/2$ on resonance. Approaching this limit is called `saturation', and there, by Eq.~(\ref{QO1}), the $\rho_{ge} \rightarrow 0$, so that the medium becomes transparent because the intense light field has effectively depolarized it. This is the basis of the SAS technique; the reduction in the resonant absorption of a weak beam (`Probe') from the reduction in polarizability of the medium caused by a stronger (`Pump') beam. 

\begin{figure}[!tbp]
\centering 
\includegraphics[width=3in]{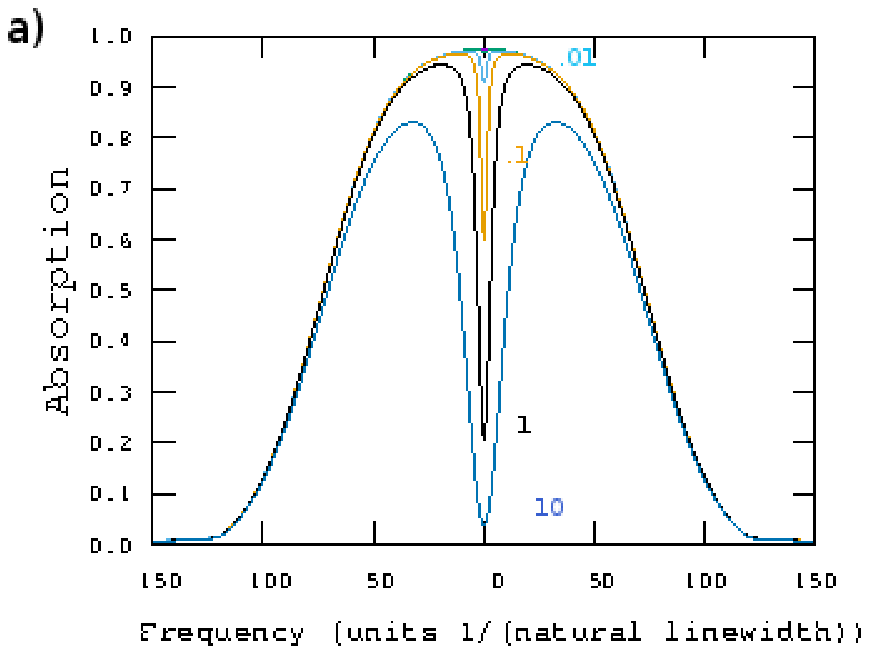}
\includegraphics[width=3in]{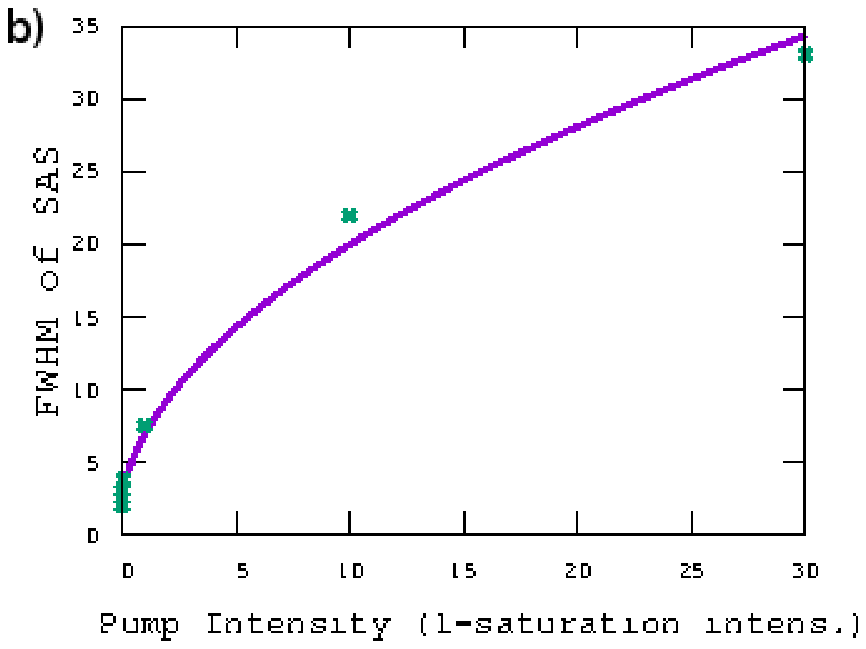}
\caption{At negligible laser spectral width and an optically thick vapor: (a) Typical absorption profile via numerical integration  for the probe beam displaying the Doppler profile (broad) and the much narrower reduction in absorption (thin) SAS feature. As the pump power is increased, the depth of the SAS feature in this simple model grows and then power broadens. Moving along from the least deep to deepest SAS feature the power increases by a factor of 10 each curve (labelled in units $I_{sat}$) (b) theory model SAS feature width in units of natural width (points) as a function of the light intensity and a two-parameter fit line based on the quadrature sum of widths, as described in text.}
\label{typicalTheorySAScurve}
\end{figure}

The simplest SAS method typically involves counterpropagating pump and probe beams derived from the same laser source.
As a consequence of the Doppler effect, atoms with a velocity component colinear to the beam then respond to the counterpropagating fields as having different frequencies. The atomic density, $n$, for each colinear velocity is proportional to the Boltzmann factor $\sim e^{-\frac{mv^2}{2k_BT}}$ (note not the Maxwell-Boltzmann distribution) where the $v$ is the component of the velocity along the propagation direction of the light beams. Thus, for atoms moving along the beams, the absorption of the weak beam (`probe') generally does not depend on the strong beam (`pump'). Atoms moving chiefly transverse to the two beams (having only a small velocity component along the beams) can be simultaneously resonant with both light fields, and so absorption of the probe beam would be reduced due to the depolarization from the saturating pump beam. Thus, convolving a flat laser frequency distribution of a prescribed width, with a Lorentzian distribution whose width is fixed by the atomic response (instrinsic width + power broadening term in denominator of Eq.~\ref{QO1}) and integrating lastly over the Boltzmann distribution to account for the atomic density in each velocity ``class", we arrive at a typical absorption profile as in Fig. \ref{typicalTheorySAScurve}a. 


For the curves in  Fig.~\ref{typicalTheorySAScurve} the natural line width of the single transition was set to 1, the Doppler width some 50 times larger.  The black curve in the graph was at a pump power of 1.0 (that is, had a Rabi frequency equal to the state's natural linewidth) and the other lines were at sequential factors of 10 above and below that value. There was assumed no angular mismatch between the pump and probe, and no transverse beam structure. In Fig.~\ref{typicalTheorySAScurve} the laser's linewidth was negligible but the vapor was optically thick.  
At intermediate powers, the spectral width of a transition is related to the quadrature sum of the widths of the two participating states. In addition to its natural linewidth, the saturating beam elicits power broadening of the transition, in order to actually get the atoms into saturation. One can think of power broadening as an effective `lifetime' of the ground state; in the presence of the optical field the ground state atoms are unstable so they `decay up' to the excited state; in other words they spend less time on average in the ground state due to absorption/stimulated emission. Achieving the saturation needed for SAS results in power broadening and indicates why even very spectrally narrow sources do not create natural line-width limited SAS resonances. Figure ~\ref{typicalTheorySAScurve}b is a plot of the simulation probe beam's SAS feature width (FWHM) as a function of pump power along with a two-parameter fit function that is the square root of a constant plus a term proportional to the power.(see Ref.[\cite{butcher}], Eq.~6.76, and Fig.~6.8 there). In a real atom like rubidium, which has  multiple ground state levels, to really understand in quantitative detail the SAS lineshape and depth it is necessary to account for the ground states' population redistribution caused by the pump field (called `optical pumping effects'). 

\begin{figure}[!tbp]
\centering 
\includegraphics[width=3in]{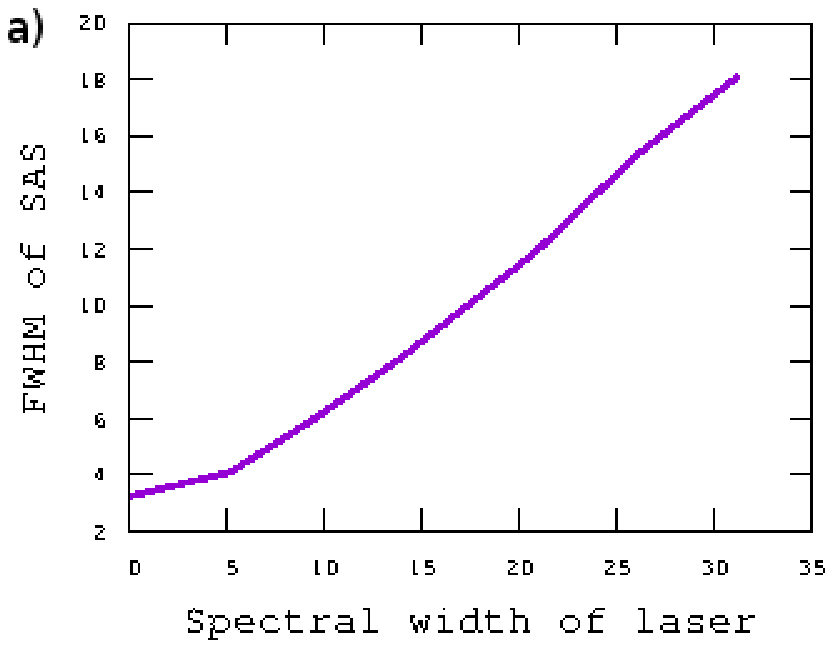}
\includegraphics[width=3in]{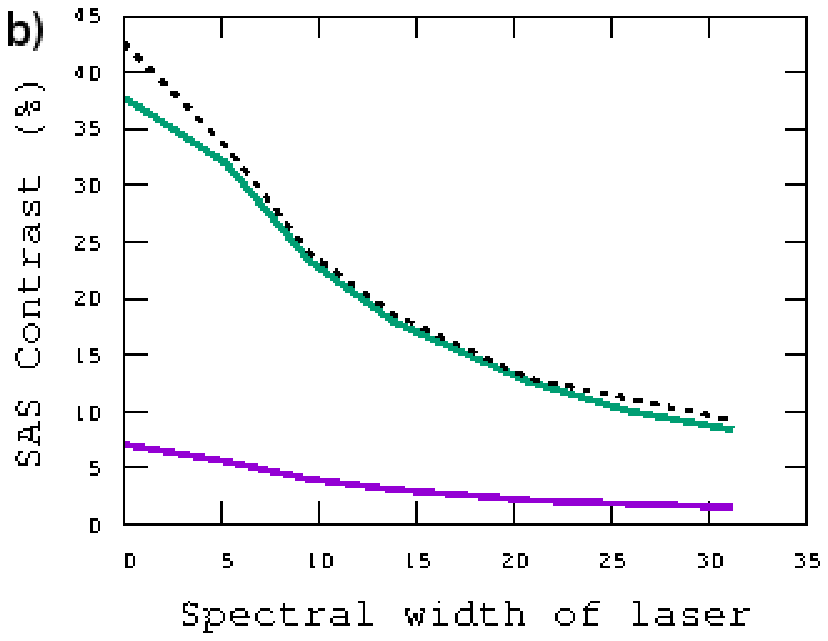}
\caption{(a) A plot of the theory-computed SAS feature width as a function of the laser linewidth (all axes in units of excited state natural line width). In this plot  $I/I_{sat}<1$.  (b) A plot of the theoretical SAS feature contrast with increasing laser linewidth. The lower line is at $I/I_{sat}$ = 0.01 whereas in the upper curve that ratio is 0.1. The dashed line is a vertically scaled  version of the lower line.}
\label{why_does_this_work}
\end{figure}

In this computer simulation of the SAS process, we can plot the SAS resonance width (FWHM) and SAS feature contrast (one minus the ratio of transparency with and without the pump field present) as a function of  the laser linewidth at fixed pump power, Fig.~\ref{why_does_this_work}. There are two important conclusions from these simulations: First, note that even at negligible laser linewidth, the FWHM of the SAS feature is a somewhat wider than natural linewidth, consistent broadly with experimental findings. Second, the laser linewidth can be as much as 10 times that of the natural line width before it starts noticeably contributing to the width of the SAS resonance, and that the SAS resonance contrast is also weakly reduced at these large laser linewidths. Although our simulation includes AC Stark terms to leading order,  there is essentially no measurable dependence of the SAS line center itself on the laser parameters. The modest dependence of the SAS feature width and contrast on laser linewidth suggests that one can accurately perform rubidium hyperfine spectroscopy with frequency-noisy free-running lasers diodes. We demonstrate that experimentally below. 

\section{Experiment:}
\begin{figure}[h!]
\centering 
\includegraphics[width=6in]{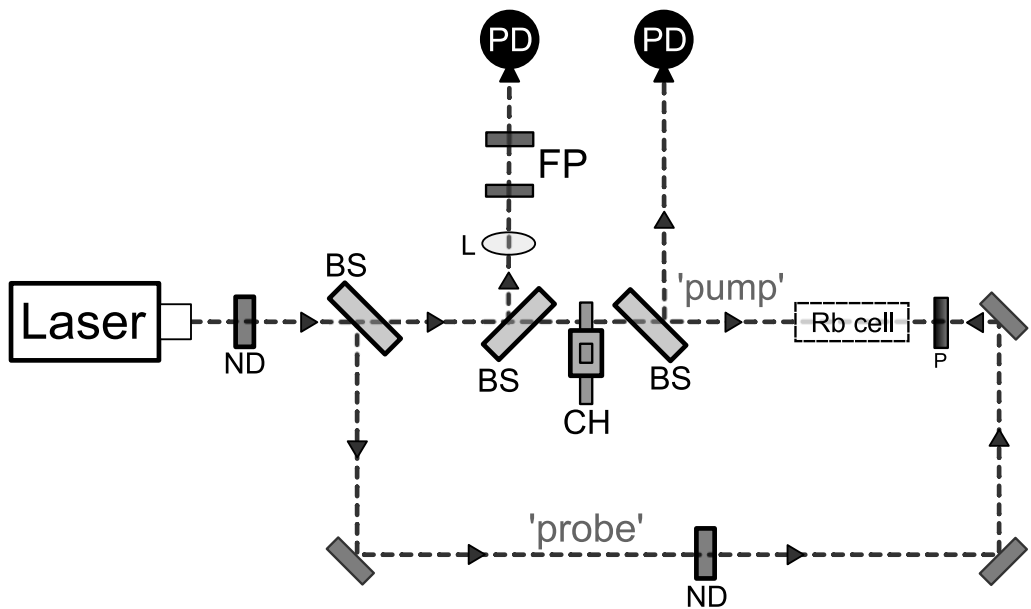}
\caption{Typical SAS setup. Rb rubidium vapor cell, FP Fabry-Perot cavity, PD photodiodes, BS non-polarizing beam splitter, P polarizer, ND neutral density filter, L lens, CH chopper for modulation transfer spectroscopy only.}
\label{SAS_setup}
\end{figure}

The experimental setup is as in Fig.~\ref{SAS_setup}.  For the instructor of the lab, once a laser source is properly tuned, lab setup and functional verification could be completed within an hour. Students will spend between one and two hours adjusting the setup (mostly beam alignment and Fabry-Perot adjustment), collecting, and completing a preliminary analysis of their data. 

As described below in reference to its contribution to the systematic error in this experiment, throughout we used a relatively low finesse ($\sim$150) aluminum-body Fabry-Perot cavity (confocal type, FSR 385MHz, 20 cm sold by Teachspin) that was not in a temperature regulated enclosure. This was a deliberate choice to make the experiment as simple and student-friendly as possible.  The chopper is not necessary for this basic experiment, but later in this manuscript we describe its use in a very simple modulation transfer protocol leading to an isotope shift measurement that is ideal for introducing students to the technique of synchronous detection. The photodectors were commodity $\sim$1 MHz,  0.3mm-square photodiodes. Commercial gold and broadband dielectric mirrors were used throughout. All the experiments described here used the same 10 cm long, 1 inch O.D. natural abundance rubidium vacuum cell in a  non-inductive, (magnetically) unshielded heated enclosure at $\sim$50$^oC$, though, at reduced signal contrast, we have also done the experiment in a room temperature cell. The polarizer was a plastic film-type suitable for NIR, useful for enhancing the SAS contrast. The neutral density filters were used both to create a weak probe beam and isolate the laser sources from back reflections (optical feedback). For each laser source the supporting electronics consisted of a commercial laser diode current driver modulated by a triangle wave from an undergraduate lab basic signal generator, along with a control loop for the laser diode temperature stabilization. For the free-running laser diodes (many were SANYO DL-7360-223H 780nm 200mW Infra-Red IR Laser Diodes, but others worked well too) we used LDC-500 (Thorlabs) current sources. Also for each of the free-running laser diodes we wrote, tested, and tuned the same simple Arduino-based temperature PID controller. Inexpensive NTC thermistors (NCP15XH103F03RC thermistors, $\sim$10KOhm at room temperature) were the sensors used in that control loop.  Each of the commercial ECDLs had additional support electronics to control/modulate the laser temperature, current and grating angle. In all experiments the pump power was between 2 and 20 milliwatts, with beam diameters ranging from 2 to 5 mm. 

\begin{figure}[h!]
\centering
\includegraphics[width=6in]{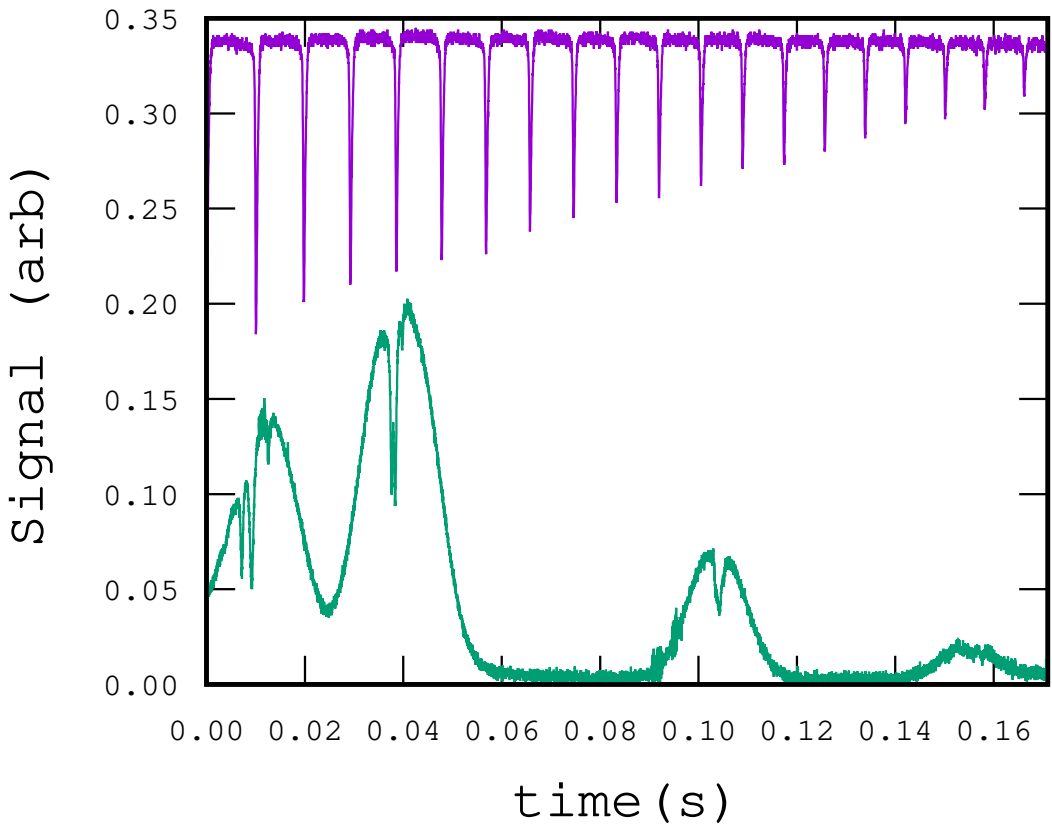}
\caption{(a) Typical D2 SAS signal (lower trace) using an ECDL, intensity ramp removed, shown with the simultaneously collected Fabry-Perot data (upper trace), displaced vertically for clarity. Note that for simplicity we have not bothered with balanced detection to remove the Doppler broadened resonances to leave just the SAS resonances.}
\label{SAS_montage}
\end{figure}

After tuning (temperature and DC current level) each laser onto resonance with the rubidium atoms, a triangle wave modulation of the current (and, for ECDL's the grating angle) on top of the DC current level scans the laser's frequency. The triangle wave parameters were adjusted so that the full hyperfine series is covered each sweep. Visually checking the regularity of the FP data indicates that the laser's frequency scan is likely to be free of frequency mode hops. The pump and probe beams are then adjusted for overlap inside the vapor cell so as to yield oscilloscope (Siglent SDS1204X-E) traces with SAS resonances on top of the absorption features, as shown in Fig.~\ref{SAS_montage}, (lower trace). The traces are stored digitally and subsequently analyzed by filling out a table of the temporal locations of each FP transmission resonance and the center of each SAS feature, using the former to locate the latter as in the D2 example provided in  Table \ref{D2example}. Best practice for doing so, as described throughout the literature, is not to fit the FP data for a single function of frequency versus sweep time, but instead to simply count FP resonances and interpolate across the nearest neighbor FP resonances.

Since both the FP and SAS resonance lineshapes in general are non-symmetric, so as to not complicate systematics, we do not fit individual lines, but simply record the location of just the peaks. This further simplifies the analysis by students. Next, to fix the frequency scale, we use the accepted value of the ground state $^{87}$Rb hyperfine splitting to determine the Free Spectral Range (FSR) of the FP cavity. Reducing the tabulated data as in
Table \ref{D2example}
to an isotope shift measurement is described in detail below,  being simple enough to do by calculator or in a brief Octave/MATLAB script. 
 
\section{Analysis/comparison} 

In Table \ref{D2example}, we include the location (by FP fringe value) for each labelled SAS resonance for a particular D2 laser. Since the D1 excited state hyperfine structure has fewer levels, it is somewhat easier to analyze and for brevity (and since the analysis is quite similar) a separate D1 example is not included here. Finally, note that under the comment column in the Table ``computed" means determined from the measured line center of the associated cross-over resonance (we take to be  
exactly half way between the contributing excited states). 

\begin{table}[h!]
\centering
\caption{Relevant SAS line centers, in units of the Fabry-Perot's Free Spectral Range (FSR). All values computed via interpolation between the nearest Fabry-Perot resonance line centers. In the comment column we delineate those which are measured outright from the data compared to those which are computed from the locations of the associated cross-over peak centers. Errors are not listed for clarity, but estimated errors are less than 10\%}
\begin{ruledtabular}
\begin{tabular}{l c c p{4cm}}
\hline	
$^{87}$Rb F=2-1' & 2.6577 & computed \\
$^{87}$Rb F=2-2' & 3.0749 & measured\\
X-over 1'+3' & 3.2306 & meas.\\
X-over 2'+3'& 3.4423 & meas. \\
$^{87}$Rb F=2-3' & 3.8034 &  meas.\\
$^{85}$Rb F=3-2' & 6.3006 & computed\\
$^{85}$Rb F=3-3' & 6.4634 & computed\\
X-over 2'+4'  & 6.5503 & meas.\\
X-over 3'+4' & 6.6317 & meas. \\
$^{85}$Rb 3-4' & 6.8000 & meas.\\
$^{85}$Rb F=2-3' &  14.560 & computed\\
$^{87}$Rb F=1-2' & 21.3019 & meas. \\
\end{tabular}
\end{ruledtabular}
\label{D2example}
\end{table}

Computing the state-averaged location of the excited hyperfine manifold for the $^{87}$Rb F=2 D2 line we have (7(3.8034)+5(3.0749)+3(2.6577)+...)/15 = 3.3314 FSR (likewise for the $^{85}$Rb F=3 in D2 light we have (9(6.8000)+7(6.4634)+5(6.3006)+...)/21 =  6.5689 FSR. The `$\dots$' refer to the small contribution we initially ignore from the lowest F' excited state.  Next, we determine the state-averaged frequency of the transition using the known F assignment and the known ground state hyperfine intervals which we take as $^{87}$Rb D2 = 6835 MHz (which from subtracting the location of the last entry from the second entry in the table we find is 18.227 FSR) and $^{85}$Rb D2 = 3036 MHz.  As explained above, we combine this with the table values to compute, for example,  the $^{87}$Rb state-averaged D2 frequency 
(5(3.3314)+3(3.3314+18.227))/8 =  10.166 FSR, whereas for the $^{85}$Rb the computation reads (7(6.5689) + 5(6.5689+18.227(3036/6835))/12 = 9.942 FSR. The difference between these two is the isotope shift in the D2 transition, for which converting to frequency via (10.166-9.942)6835/18.227 =  84 MHz (compare with 81 MHz of  Ref.~[$^2$] and the accepted value of  77 MHz of Ref.~[\cite{beacham, gibbs, aldridge}]). 

The normal mass shift (NMS) alone for this transition (780.25nm photon's  f= 3.84 x 10$^{14}$ Hz) is found by $\mu$-scaling, using $df/f  = |d\lambda/\lambda|$ leading to $df$ = 57.13 MHz. That means that the  84-57 =   27 MHz is apparently the  “nuclear”  part (NS) of the isotope shift between $^{85}$Rb and $^{87}$Rb. The best fit value for this in the literature is about 20 MHz.\cite{aldridge} 

Now assuming that whole NS part is due to the ground S state (since those electrons spend so much more time than non-S electrons inside the nucleus) only, we can recast the measured isotope shift into a ground state energy difference between the isotopes, relative to the continuum. Such a difference indicates a chemical difference between different isotopes, for one example, responsible for the fact that the binding energy of deuterium to a typical molecule is larger than that of hydrogen. This leads to the spontaneous segregation into more heavily deuterated species and changes in the chemical reaction rates, some of which have profound physiological consequences, including toxicity of D$_2$O. Of course, as 
discussed earlier, part of the overall electron binding energy scales with $\mu$, so that one can expect the ratio of the normal part of the isotope shift (NS) to the isotope energy difference be the same as the ratio of the optical excitation energy to the known ionization energy of Rb (4.184 eV). Thus  using the fact that our D2 transition at 780.25 nm has an energy of 1.59 eV, we get 57.13(4.184/1.59)+27 = 177 MHz as the ground state energy shift between $^{85}$Rb and $^{87}$Rb relative to the continuum. 

Multiple runs with different sources and at either wavelength (D1 or D2) indicates that the error in this ground state energy shift can be as large as $\pm$ 15 MHz. The accepted value for the ground state energy shift is 164 MHz.\cite{aldridge} The difference from our result detailed here is chiefly a systematic error from terms we ignored in computing the ``center of mass" of the excited state manifolds (those `$\ldots$' in the above). Note that if one includes as a naive estimate for the missing transitions $^{85}$Rb F=3 → F'=1 and the $^{87}$Rb F=2 → F'=0  contributing to the `$\ldots$' as being as far in energy from their neighbor (the  $^{85}$Rb F=3 → F'=2 and the $^{87}$Rb F=2 → F'=1) as those states are from their respective neighbors (the $^{85}$Rb F=3 → F'=3 and the $^{87}$Rb F=2 → F'=2), one finds an isotope shift of 79 MHz (resulting in a ground state energy shift of  172 MHz), somewhat closer to the accepted value.

\begin{figure}[h!]
\centering
\includegraphics[width=3in]{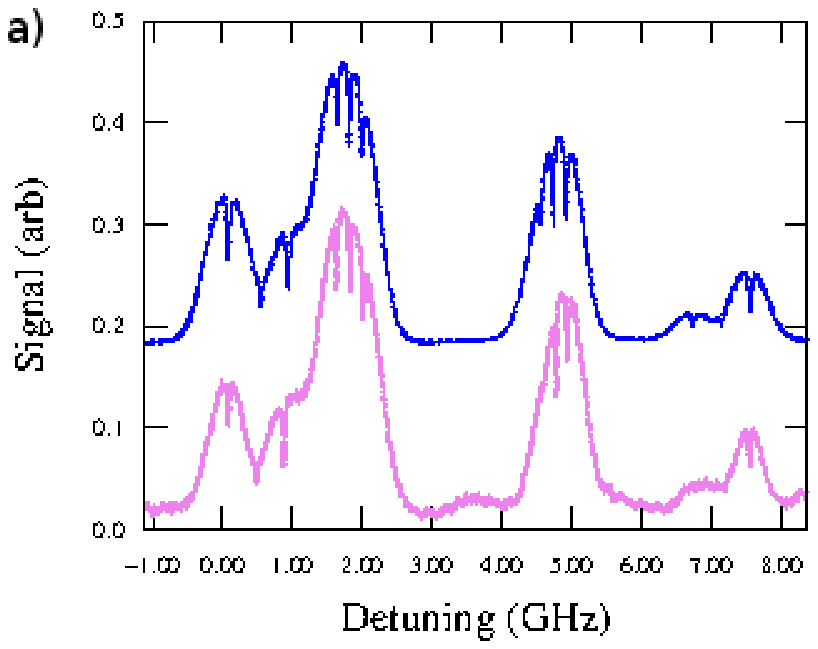}
\includegraphics[width=3in]{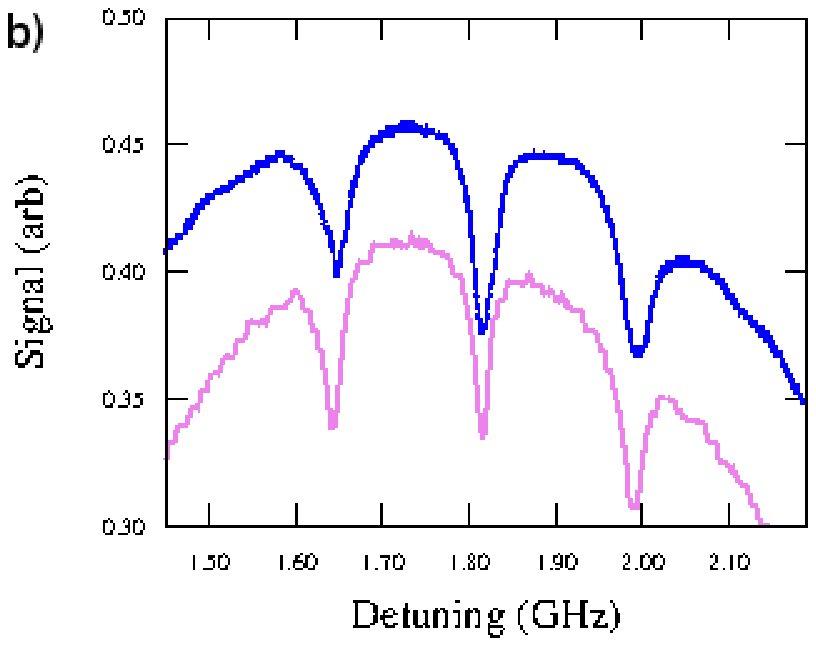}
\caption{(a) D1 SAS spectra from a free running diode laser (light trace) vertically offset for clarity from that using a commercial (Vortex\textsuperscript{TM}, New Focus) ECDL (dark trace). (b) Detail of the overlay of the F=2 $^{85}$Rb SAS resonances showing comparable SAS resonance width and contrast for some of the lines.}
\label{direct_compare}
\end{figure}

Figure~\ref{direct_compare} is a direct comparison of two different D1 SAS traces, one with a free-running laser diode and one with a commercial (Vortex\textsuperscript{TM} new Focus) ECDL on the same frequency scale. Beyond just visual comparison of the signals, an isotope shift measurement yields a more detailed quantitative comparison of the spectroscopic data obtained from the various light sources. 



\begin{table}[h!]
\centering
\caption{Comparison of measured isotope shifts (MHz) using various sources. The last column is the computed ground state hyperfine interval ratio between the two isotopes present, and in row 1 are the best current values of these quantities from the literature. All computed isotope shifts $\pm$10 MHz and the `$^*$'' is for the data collected for Fig.~\ref{SAS_montage} but without including the unresolved small F' (excited) states estimated location...see text.} 
\begin{ruledtabular}
\begin{tabular}{l c c c c p{5cm}}
\hline	
Ref.[\cite{aldridge}]  & - & 78 & 2.2514\\
Free Running \#1 & D1 & 69.7 &  2.2468 \\
Free Running \#1'& D1 & 71.4 &  2.2498 \\
Free Running \#2 & D1 & 73.1 & 2.2457 \\
Free Running \#3 & D1 & 73.4 & 2.2403 \\
Free Running \#4 & D2 & 79.4 & 2.2581 \\
ECDL Vortex\textsuperscript{TM} (New Focus) & D1 & 80.3 & 2.2487 \\
ECDL DLS\textsuperscript{TM} (Teachspin)$^*$ & D2 & 84.2 &  2.2553 \\
ECDL Vortex\textsuperscript{TM} (via MTS) & D1 & 74.1 &  2.2555\\
\end{tabular}
\end{ruledtabular}
\label{summaryLasers}
\end{table}

The isotope shift measurements by the method above using both D1 and D2 free-running laser diodes and commercial ECDLs is summarized in Table~\ref{summaryLasers}.  All of these measurements of the isotope shift are within a few MHz (about 10\%) of each other and of the accepted value (78MHz, \cite{aldridge}).  Although we have performed this experiment on only a few lasers of each type, we expect these results to be typical. We conclude from this comparison that using a free-running laser diode to interrogate warm atoms yields spectroscopic data of essentially identical quality to that of ECDLs. In recognition of their comparatively simpler construction, use and much lower cost,  free-running laser diodes for rubidium spectroscopy are practical and advantageous for these undergraduate laboratory experiments. 

\section{Modulation Transfer Spectroscopy use in Isotope Shift Measurement}

\begin{figure}[h!]
\centering
\includegraphics[width=6in]{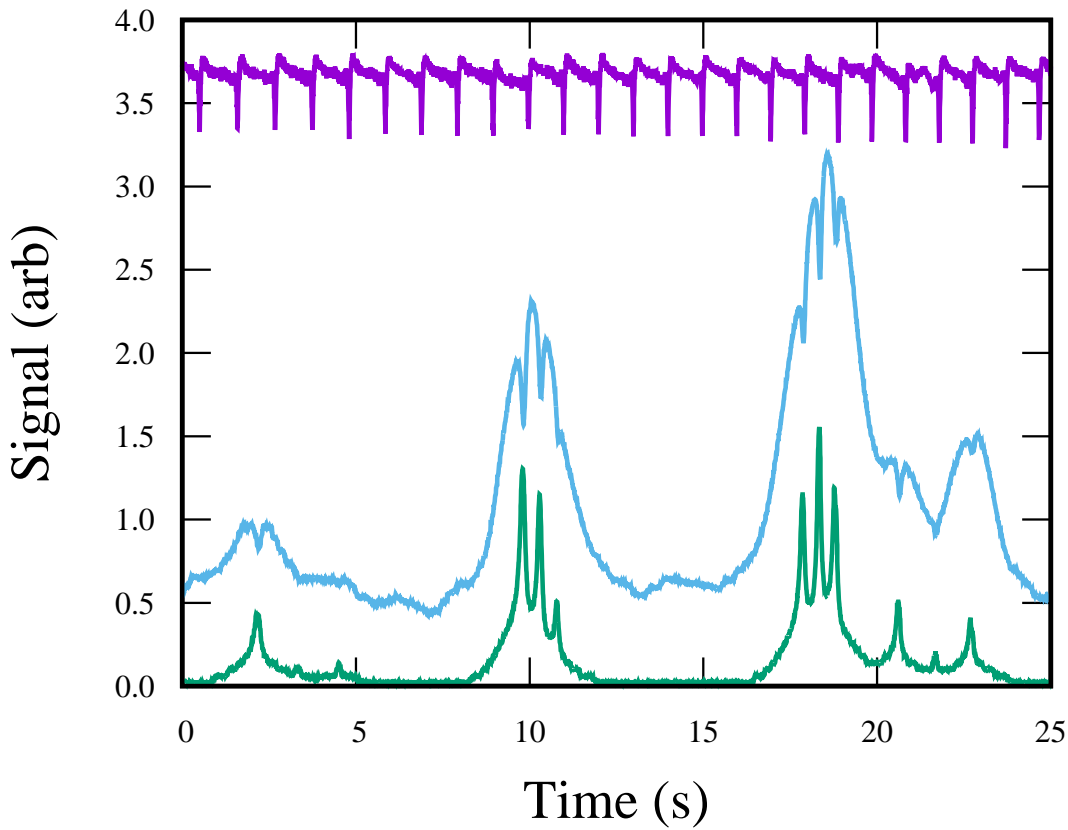}
\caption{Modulation transfer experiment lock-in output (lower trace) overlayed on the D1 SAS signal (middle trace, vertically offset) with Fabry-Perot resonances (uppermost trace). A 3100 Hz chop was used and the gain on the chopped pump PD was increased to get at least 0.5 Vpp for the SRS830 lock-in amplifier reference input. Other lock-in parameters were 1ms time constant, 100mV scale.}
\label{MTS_graphic}
\end{figure}

Because it is a two-beam non-linear optical process, 
SAS lends itself to a student-friendly demonstration of the utility of synchronous signal detection via modulation transfer spectroscopy.  One only need turn on the chopper (shown in Fig.~\ref{SAS_setup})  and correlate the pump beam interruptions with the probe field's intensity variations.\cite{schmitt} We use as the reference input to the lock-in amplifier the voltage from an additional amplified photodetector (not shown in Fig.~\ref{SAS_setup}) that intercepts a part of the pump beam before the cell. The probe field's transmission photodetector voltage is input to the lock-in. The laser scan is then slowed (to about a 25 second period) and the SAS resonances appear as peaks in the amplifier's output as shown in Fig.~\ref{MTS_graphic}. For those data the 795nm New Focus Vortex\textsuperscript{TM} ECDL laser was used, and analysis of these lock-in data yields an isotope shift of 74 MHz and the measured ground state hyperfine splitting ratio between isotopes that is only about 0.2\% higher than the accepted value. 

Although there are countless other ways of displaying the utility of lock-in detection in SAS and other non-linear optical experiments, this one is conceptually simple, and ideal for a student introduction to the technique. Of course, for this to work well students have to take care to eliminate any extraneous pump light from entering the probe transmission photodiode. Since these optical fields are counterpropagating this is usually easy to arrange.

\section{Conclusion} 
Finally, we address several factors that complicate the interpretation of these SAS data as a measurement of an isotope shift. Some of the confounding effects include displacements of the SAS resonances due to AC stark shifts, optical pumping effects that may disturb the SAS resonance line center,  residual magnetic field shifts, line distortions from AC response of detectors during the sweep (typically at about 20-40 Hz), laser+Fabry-Perot cavity frequency and intensity drift during sweep, effects of transverse beam intensity profiles and their spatial overlap, Fabry-Perot cavity non-linearity, and finally, ubiquitous table (mostly pointing)/electrical noise. Although most of these effects can deleteriously `pull' the line center one way or another, our crude approach of locating the peaks probably leads to the  largest contribution to the error in the determination of the isotope shift. 

We have measured some of these effects, including the short term intrinsic spectral stability of the 3d printed plastic laser head plus custom Arduino-based digital control loop we use (to be described in a forthcoming publication) for our free-running laser diode sources, so as to compare it to commercially available products. By parking the lasers at either an edge of an absorption line or using a pair of invar confocal cavities (Thorlabs SA200-8B, FSR 1.5 GHz) and  recording  the  transmission  changes  over  time  we  have  measured  RMS short term  intrinsic laser frequency drifts of typically less than 17 MHz/s for the 3d printed plastic laser head (795nm) and Arduino temperature stabilization. This is to be compared with less than 3 MHz/s in the Newfocus Vortex\textsuperscript{TM} laser (also at 795nm).  Of course, since one  captures  both the SAS resonance lines and the FP resonance lines during the same (single) trace, the limitation here is primarily on the FP stability. For ease of student use in all the experiments described here we used an FP (Teachspin, FSR 380MHz, 20 cm, no temperature stabilization) with an aluminum body, which we estimate may drifted as fast as 60 MHz/s, leading to line center errors not more than a few MHz level in a single trace.

On the basis of the overall errors in the various attempts, we surmise that the forgoing effects combine randomly at this level. Exploring which of these other effects may dominate the systematic errors of this measurement is an important topic under investigation. 
Frequency-noisy free-running laser diodes are ideal for a straightforward, pedagogically useful and inexpensive undergraduate advanced laboratory measurement of the isotope shift in rubidium. This highlights and extends access to this non-linear optical experiment that reveals subtleties of atomic structure and can refine student's microphysical `picture' of the atom, most notably, by challenging common misconceptions about the motion of the electron in the ground state. In computing the isotope shift, students also get a glimpse of the role of nuclear mass and size in electronic spectra. Beyond a demonstration of the use of free-running laser diodes with SAS, we evaluated a straightforward quantum optics based theory model that explains why the SAS resonances are somewhat sub-laser linewidth.  Because of its rich utility, relative setup ease and low cost, it is hoped that this study will aid the diffusion of this laboratory across additional undergraduate advanced laboratories.



\begin{acknowledgments} It is a pleasure to acknowledge insightful discussions with S. Bali, E. Hazlett and D. Van Baak regarding this work. This work was supported in part by the U.S. National Science Foundation (Grant DMR-1609077) and the Advanced Manufacturing Research Center at YSU.  
\end{acknowledgments}

\end{document}